\documentclass{webofc}\usepackage[varg]{txfonts}

\usepackage{epsfig,graphics,amssymb,amsmath,mathrsfs,color,colortbl,
bm,booktabs,cool,dcolumn}
\begin{document}\title{Multiquark-Oriented QCD Sum Rules}
\author{Wolfgang Lucha\inst{1}\fnsep\thanks{\email
{Wolfgang.Lucha@oeaw.ac.at}}\and Dmitri Melikhov\inst{2,3,4}\fnsep
\thanks{\email{dmitri_melikhov@gmx.de}}\and Hagop Sazdjian\inst{5}
\fnsep\thanks{\email{sazdjian@ijclab.in2p3.fr}}}\institute{Institute
for High Energy Physics, Austrian Academy of Sciences,
Nikolsdorfergasse 18,\\A-1050 Vienna, Austria\and D.~V.~Skobeltsyn
Institute of Nuclear Physics, M.~V.~Lomonosov Moscow State
University,\\119991 Moscow, Russia\and Joint Institute for Nuclear
Research, 141980 Dubna, Russia\and Faculty of Physics, University
of Vienna, Boltzmanngasse 5, A-1090 Vienna, Austria\and
Universit\'e Paris-Saclay, CNRS/IN2P3, IJCLab, 91405 Orsay,
France}\abstract{We propose to increase the factual reliability of
descriptions of exotic multiquark hadrons utilizing the approach
to bound states of strongly interacting constituents known as QCD
sum rules, by allowing exclusively all contributions that
potentially bear some relevance for multiquark states to enter the
correlation functions that form the main ingredient of this
framework. The route to this goal is illustrated for the
(presumably least involved) special case of tetraquark~states.}
\maketitle

\section{Goal: Adaptation of QCD Sum-Rule Approach to Multiquark
States}Evaluation of \emph{correlation functions\/} of operators
interpolating one's hadrons of interest at both phenomenological
(hadron) and fundamental (QCD) levels by application of operator
product expansion \cite{KGW} and Borel transformations, and
assuming a complete cancellation of hadron~and perturbative-QCD
contributions above optimized \cite{ET1,ET2,ET3} effective
thresholds, provides analytic relationships, denoted as \emph{QCD
sum rules\/} \cite{QSR}, between, on the one hand, observable
properties of hadrons and, on the other hand, basic parameters of
QCD. In this context QCD accounts for both perturbative
contributions, given by integrals over spectral densities, and
nonperturbative power corrections (that involve vacuum condensates
multiplied by powers of Borel variables).

The colour-singlet bound states governed by QCD encompass not just
all quark--antiquark mesons and three-quark baryons, labelled
\emph{conventional\/} hadrons, but also \emph{multiquark\/}
hadrons: tetraquarks, pentaquarks, hexaquarks, etc. For attempts
to take hold of such exotic hadrons by QCD sum rules, a systematic
scrutiny \cite{ESRp,ESRr,TMA1,TMA2,TMA3} points to the necessity
of adapting the traditional wisdom to the peculiarities of
multiquarks: Retaining just contributions potentially capable of
providing information on multiquark features by getting rid of all
contributions not related to multiquarks yields the novel sort of
\emph{multiquark-adequate\/} QCD sum rules; for some tetraquark
categories \cite{LMS4}, all of the latter involve, at least, two
gluon exchanges of appropriate topology.

\section{Proof of Assertion by Brief Sketch of Underlying Line of
Argument}We illustrate our considerations for the simplest type of
multiquark hadrons, the \emph{tetraquark\/} $T$, a meson
constituted of two antiquarks $\overline q_a,\overline q_c$ and
two quarks $q_b,q_d$ (of masses $m_a,m_b,m_c,m_d$):\pagebreak
$$T=[\overline q_a\,q_b\,\overline q_c\,q_d]\ ,\qquad
a,b,c,d\in\{u,d,s,c,b\}\ .$$Leaving aside the spin degrees of
freedom, by application of a Fierz transformation \cite{MF} every
\emph{tetraquark-interpolating operator\/} may be proven
\cite{RLJ} to be equivalent to a linear combination of just two
products $\theta$ of colour-singlet quark--antiquark bilinear
currents $j_{\overline ab}(x)\equiv\overline q_a(x)\,q_b(x)$:
$$\theta_{\overline ab\overline cd}(x)\equiv j_{\overline
ab}(x)\,j_{\overline cd}(x)\ ,\qquad\theta_{\overline ad\overline
cb}(x)\equiv j_{\overline ad}(x)\,j_{\overline cb}(x)\ .$$

Accordingly, we regard it preferable to exploit (potential)
manifestations of tetraquarks by means of intermediate-state poles
in the scattering amplitudes of two conventional mesons (of
momenta $p_1,p_2$) to two conventional mesons, by scrutiny of
vacuum averages of time-ordered products of four quark--antiquark
bilinear currents $j^{(\dag)}$, i.e., of \emph{four-point
correlation~functions\/}$$\left\langle{\rm T}\!\left(j(y)\,j(y')\,
j^\dag(x)\,j^\dag(x')\right)\right\rangle.$$Configuration-space
contractions of the latter yield all required types of correlation
functions:\begin{itemize}\item two-point correlation functions of
two operators $\theta$, interpolating the tetraquark investigated,
$$\left\langle{\rm T}\!\left(\theta(y)\,\theta^\dag(x)\right)
\right\rangle=\lim_{\underset{\scriptstyle y'\to y}{\scriptstyle
x'\to x}}\left\langle{\rm T}\!\left(j(y)\,j(y')\,j^\dag(x)\,
j^\dag(x')\right)\right\rangle;$$\item three-point correlators of
one current $\theta$ and two conventional-meson interpolating
currents~$j$$$\left\langle{\rm T}\!\left(j(y)\,j(y')\,
\theta^\dag(x)\right)\right\rangle=\lim_{x'\to x}\left\langle{\rm
T}\!\left(j(y)\,j(y')\,j^\dag(x)\,j^\dag(x')\right)\right\rangle.$$
\end{itemize}

Next, we need to formulate an exact and easily applicable
\emph{criterion\/} \cite{TQC1,TQC2} that allows for unequivocal
identification of all contributions impacting any correlation
function under~study:\begin{description}\item[Tetraquark-phile]
\cite{TQP1,TQP2} contributions depend in a \emph{nonpolynomial\/}
form on the Mandelstam variable $s\equiv(p_1+p_2)^2$ and -- as can
be established by fulfilment of the Landau equations \cite{LDL} --
enable true intermediate four-quark states, by exhibiting branch
cuts that start at \emph{branch~points\/}$$\hat
s=(m_a+m_b+m_c+m_d)^2\ .$$\end{description}

As an example, we focus to the subset of \emph{flavour-exotic\/}
tetraquarks, which encompass two quarks and two antiquarks
carrying four \emph{unequal\/} quark flavours, generically
labelled~$a,b,c,d$.

For such choice of quark-flavour contents, in the course of
analysis it proves advantageous to discriminate two categories of
four-point correlation functions, differing in the distributions
of the four quark flavours among the pairs of interpolating
currents $j^{(\dag)}$ in initial or final states:\begin{itemize}
\item\emph{flavour-retaining\/} correlation functions
\cite{ESRp,TMA1,TMA2}, identified by identical flavour
distributions,$$\left\langle{\rm T}\!\left(j_{\overline ab}(y)\,
j_{\overline cd}(y')\,j^\dag_{\overline ab}(x)\,j^\dag_{\overline
cd}(x')\right)\right\rangle,\qquad\left\langle{\rm T}\!\left(
j_{\overline ad}(y)\,j_{\overline cb}(y')\,j^\dag_{\overline
ad}(x)\,j^\dag_{\overline cb}(x')\right)\right\rangle;$$
\item\emph{flavour-rearranging\/} correlation functions
\cite{ESRp,ESRr,TMA1,TMA2}, identified by unequal flavour
distributions,$$\left\langle{\rm T}\!\left(j_{\overline
ab}(y)\,j_{\overline cd}(y')\,j^\dag_{\overline ad}(x)\,
j^\dag_{\overline cb}(x')\right)\right\rangle.$$\end{itemize}

\begin{figure}[ht]\centering{\includegraphics[scale=.30696,clip]
{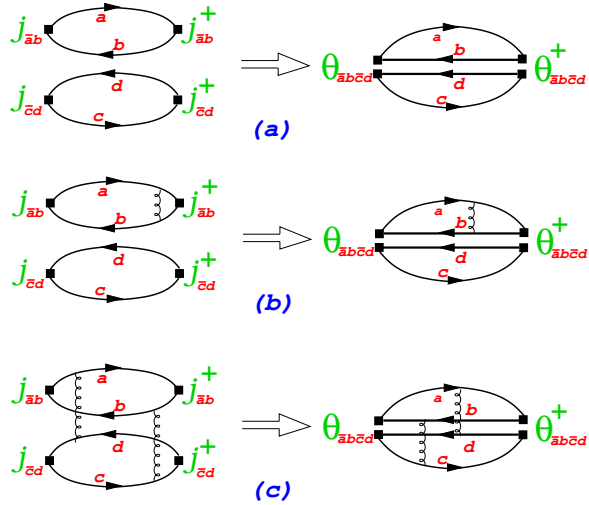}}\caption{Flavour-preserving correlation
functions of two tetraquark-interpolating operators $\theta$,
resulting from double configuration-space contractions of a pair
of operators $j$ interpolating conventional mesons: generic
contributions of lowest (a), next-to-lowest (b) and
next-to-next-to-lowest (c) perturbative orders.}\label{1a}
\end{figure}

\begin{figure}[hb]\centering{\includegraphics[scale=.30696,clip]
{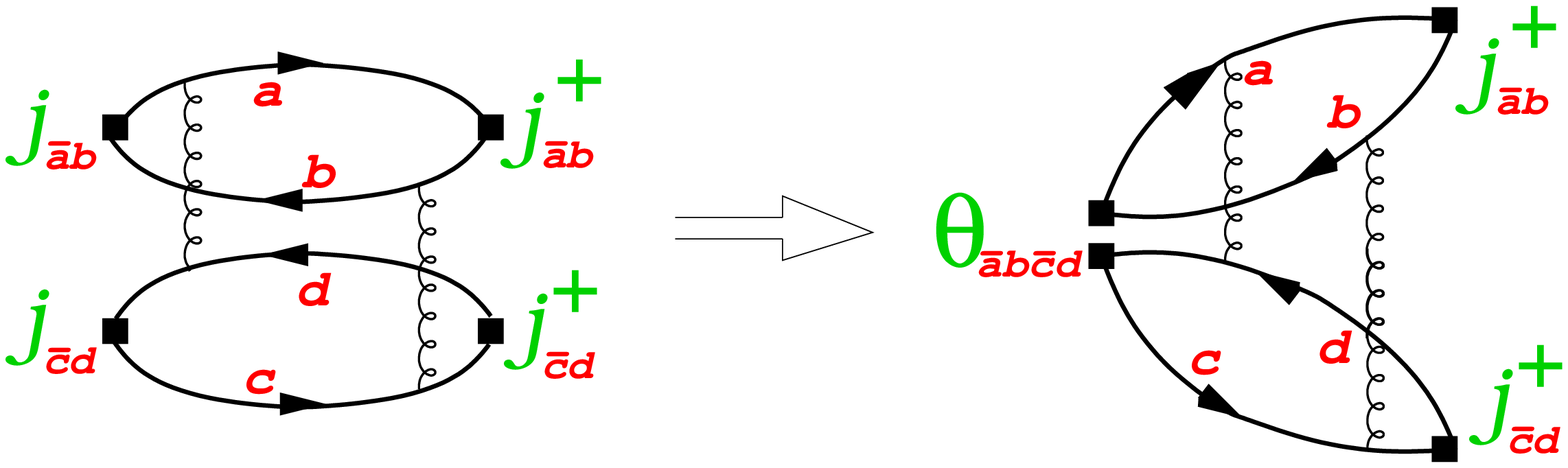}\\{\sl\textcolor{blue}{(a)}}\\
\includegraphics[scale=.30696,clip]{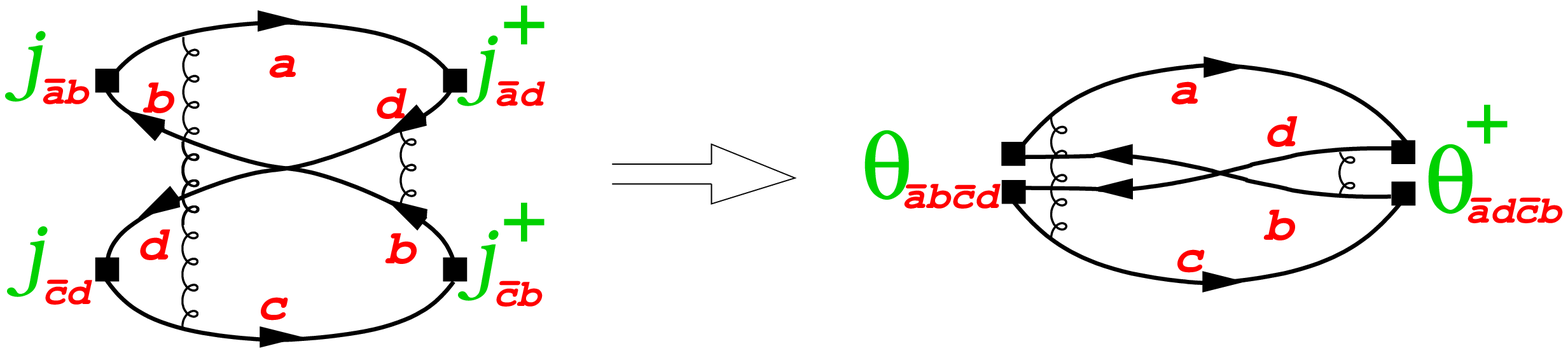}\\
{\sl\textcolor{blue}{(b)}}\\\includegraphics[scale=.30696,clip]
{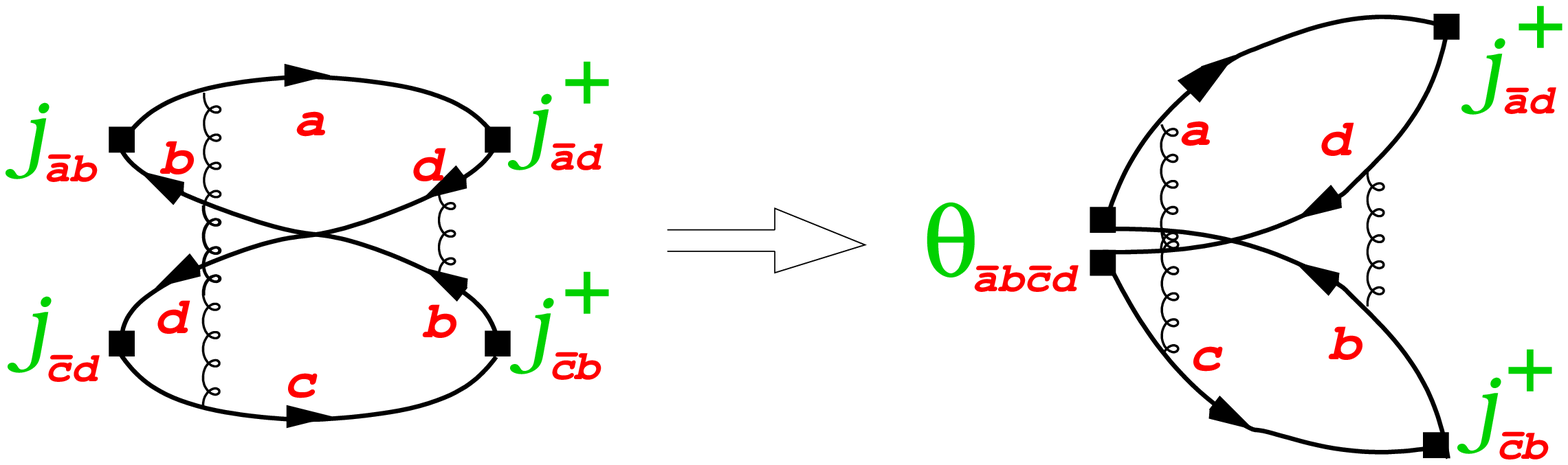}}\\{\sl\textcolor{blue}{(c)}}\caption
{Exemplary \emph{tetraquark-phile\/} contributions of lowest
perturbative order to (a) \emph{flavour-retaining\/} correlation
functions of one tetraquark-interpolating current $\theta$ and two
quark--antiquark bilinear currents $j$ and to
\emph{flavour-reordering\/} correlation functions of either (b) a
pair of tetraquark-interpolating operators $\theta$ or (c) one
tetraquark-interpolating operator $\theta$ and two ordinary-meson
interpolating currents $j$, inferred by single (a,c) or double (b)
configuration-space contraction of two meson-interpolating
currents $j$, resp.}\label{1b}\end{figure}

Armed with our rigorous criterion \cite{TQC1,TQC2} above, for
every two- or three-point correlation function it is then a
straightforward task to identify the tetraquark-phile
\cite{TQP1,TQP2} contributions:\begin{description}
\item[Flavour-retaining ones] \cite{ESRp,TMA1,TMA2,TMA3} with no
or at most a single gluon exchange [Fig.~\ref{1a}(a,b)] are
spatially separable and vanish at all or involve $s$ in merely
polynomial form, implying~that just ones with two or more relevant
gluon exchanges do affect tetraquarks [Fig.~\ref{1a}(c)
and~Fig.~\ref{1b}(a)].\item[Flavour-reordering ones]
\cite{ESRp,ESRr,TMA1,TMA2,TMA3} are not spatially separable and
will support tetraquarks only if resorting to two or more gluon
exchanges, as revealed by the Landau equations
[Fig.~\ref{1b}(b,c)].\end{description}\pagebreak


\section{Outcome: QCD Sum Rules Tailored to Peculiarities of
Tetraquarks}
\begin{figure}[ht]\centering{\includegraphics[scale=.27765,clip]
{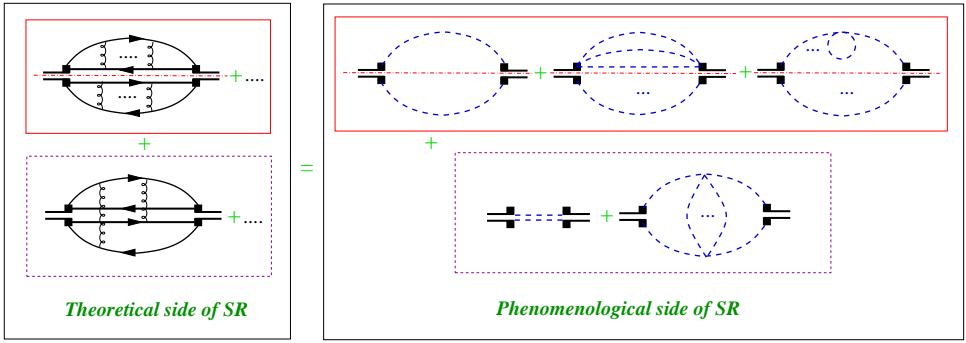}}\caption{Decomposing a generic
\emph{flavour-retaining\/} relationship into a pair of
\emph{conventional\/}-meson QCD sum rules (top row) and a QCD sum
rule that possibly involves, among others, tetraquarks (bottom
row).}\label{2}\end{figure}

\begin{figure}[ht]\centering{\includegraphics[scale=.372444,clip]
{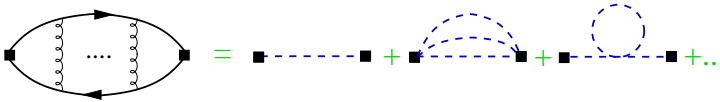}}\caption{Generic QCD sum rule for
\emph{ordinary\/} mesons (blue dashed lines) following traditional
folklore.}\label{3}\end{figure}

\begin{figure}[ht]\centering{\includegraphics[scale=.372444,clip]
{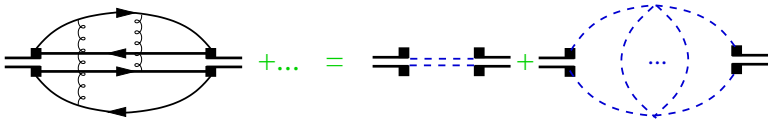}}\caption{Tetraquark-adequate QCD sum rule,
possibly involving tetraquarks (blue dashed double line).}
\label{4}\end{figure}

\begin{figure}[ht]\centering{\includegraphics[scale=.30696,clip]
{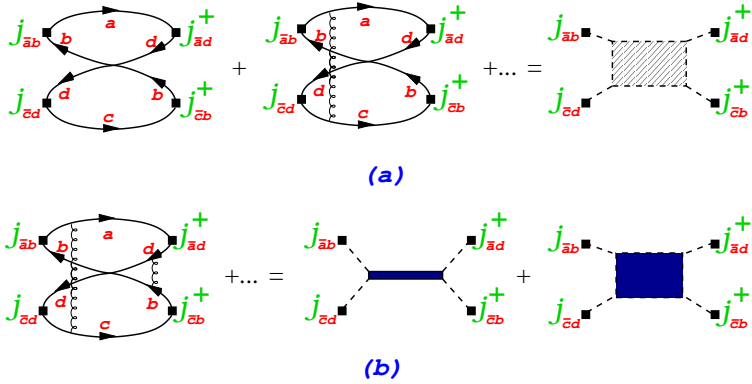}}\caption{Decomposing each
\emph{flavour-rearranging\/} relation into (a) one without
four-quark or two-meson $s$-channel cut and (b) a
\emph{tetraquark-adequate\/} QCD sum rule supporting tetraquarks
(blue horizontal bar).}\label{5}\end{figure}

Upon feeding the above concepts or insights into the
traditional framework of QCD sum rules, it becomes feasible to
disentangle, in any deduced relation, the unavoidable
intertwinement of expressions genuinely related to tetraquarks and
those involving exclusively ordinary mesons:\begin{itemize}
\item Every flavour-preserving relation (Fig.~\ref{2}) consists of
two kinds of QCD sum rules \cite{ESRp,TMA1,TMA2,TMA3}; the first
of these provides two unconnected QCD sum rules for conventional
mesons~(Fig.~\ref{3}) and the second one the required
\emph{tetraquark-adequate\/} QCD sum rule (Fig.~\ref{4})
\cite{ESRp}. Refraining from one contraction of interpolating
currents $j^{(\dag)}$ disentangles the three-point analogue
\cite{ESRp}.\item By investing a little bit more efforts, every
flavour-rearranging relation may be unscrambled
\cite{ESRp,ESRr,TMA1,TMA2,TMA3} and systematically reorganized as
two relationships, one that \emph{does not\/} [Fig.~\ref{5}(a)]
and one that \emph{does\/} [Fig.~\ref{5}(b)] involve two-meson
$s$-channel cuts and potentially tetraquark~poles.\end{itemize}


In addition to the perturbative contributions to the QCD sum rules
aimed at, various power corrections have to be considered, too.
Their inclusion is more tricky but accomplishable \cite{LMS10}.

Ultimately, incorporating the aforementioned findings should offer
a \emph{reliable\/} extraction of the fundamental properties of
tetraquarks [their masses $M$, decay constants $f_{\overline
ab\overline cd}\equiv\langle0|\theta_{\overline ab\overline
cd}|T\rangle$ and $f_{\overline ad\overline cb}\equiv\langle0|
\theta_{\overline ad\overline cb}|T\rangle$, and momentum-space
amplitudes $A(T\to j_{\overline ab}\,j_{\overline cd})$ and
$A(T\to j_{\overline ad}\,j_{\overline cb})$] from
\emph{tetraquark-adequate QCD sum rules\/} for all two- and
three-point correlation functions~of\newpage\noindent interest
involving flavour-preserving or -reordering tetraquark-phile
spectral densities $\rho_{\rm p,r}$~and $\Delta_{\rm p,r}$ and
effective thresholds $s_{\rm eff}(\tau)$ depending on Borel
parameters $\tau$, symbolically of the form\begin{align*}&
(f_{\overline ab\overline cd})^2\exp(-M^2\,\tau)\\&=
\int\limits_{\hat s}^{s_{\rm eff}(\tau)}{\rm d}s\exp(-s\,\tau)
\,\rho_{\rm p}(s)+\mbox{Borel-transformed power corrections}\
,\\&f_{\overline ab\overline cd}\,A(T\to j_{\overline ab}\,
j_{\overline cd})\exp(-M^2\,\tau)\\&=\int\limits_{\hat s}^{s_{\rm
eff}(\tau)}{\rm d}s\exp(-s\,\tau)\,\Delta_{\rm p}(s)+
\mbox{Borel-transformed power corrections}\ ,\\&f_{\overline
ab\overline cd}\,f_{\overline ad\overline cb}\exp(-M^2\,\tau)\\&=
\int\limits_{\hat s}^{s_{\rm eff}(\tau)}{\rm d}s\exp(-s\,\tau)\,
\rho_{\rm r}(s)+\mbox{Borel-transformed power corrections}\ ,\\&
f_{\overline ad\overline cb}\,A(T\to j_{\overline ab}\,
j_{\overline cd})\exp(-M^2\,\tau)\\&=\int\limits_{\hat s}^{s_{\rm
eff}(\tau)}{\rm d}s\exp(-s\,\tau)\,\Delta_{\rm r}(s)+
\mbox{Borel-transformed power corrections}\ .\end{align*}

Last but not least, the \emph{cluster reducibility\/} \cite{LMS7}
of any multiquark (its group-theory-implied potential
decomposition into a cluster of ordinary hadrons) requires to
regard a multiquark,~on an equal footing, as tightly bound compact
and loosely bound molecular-type
hadron~\mbox{\cite{LMS111,LMS112}}.\vspace{3.39778ex}

\begin{acknowledgement}\noindent{\bf Acknowledgements.~~}D.~M.\ and
H.~S.\ express sincere gratitude for support by joint CNRS/RFBR
Grant PRC Russia/19-52-15022, D.~M.\ for support by the Austrian
Science Fund (FWF), Project P29028-N27, H.~S.\ for support by EU
research and innovation program Horizon 2020 under Grant
Agreement~824093.\end{acknowledgement}

\end{document}